# Les Agents comme des interpréteurs Scheme :
# Spécification dynamique par la communication

# Agents as Scheme Interpreters:
# Enabling Dynamic Specification by Communicating

Clement Jonquet    Stefano A. Cerri

LIRMM - Université Montpellier II
161, rue Ada
34392 Montpellier Cedex 5 - France

{cerri, jonquet}@lirmm.fr

## Résumé

*Nous avons proposé dans de précédents papiers une extension et une implémentation du modèle STROBE, qui considère les Agents comme des interpréteurs Scheme. Ces Agents sont capables d'interpréter des messages dans des environnements donnés incluant un interpréteur qui apprend de la conversation et donc qui représente l'évolution de sa connaissance au niveau méta. Quand ces interpréteurs sont non déterministes, le dialogue consiste à raffiner les spécifications d'un problème par des ensembles de contraintes. Ce papier présente un exemple de génération dynamique de service – tels qu'ils sont nécessaires sur le GRID – exploitant des Agents STROBE équipés d'un interpréteur non déterministe. Il montre comment réaliser la spécification dynamique d'un problème. Puis il illustre comment ces principes peuvent être intéressants pour d'autres applications. Les détails de l'implémentation ne sont pas fournis ici mais sont disponibles.*

## Mots Clef

Communication Agent, interprétation dynamique de message, évaluation non déterministe, STROBE, spécification dynamique, contraintes, dialogue.

## Abstract

*We proposed in previous papers an extension and an implementation of the STROBE model, which regards the Agents as Scheme interpreters. These Agents are able to interpret messages in a dedicated environment including an interpreter that learns from the current conversation therefore representing evolving meta-level Agent knowledge. When the Agent interpreter is a nondeterministic one, the dialogues may consist of subsequent refinements of specifications in the form of constraint sets. The paper presents a worked out example of dynamic service generation – such as necessary on Grid – by exploiting STROBE Agents equipped with a nondeterministic interpreter. It shows how enabling dynamic specification of a problem. Then, it illustrates how these principles could be effective for other applications. Details of the implementation are not provided here, but are available.*

## 1. Introduction

La transmission du savoir est quelque chose d'essentiel pour toutes les sociétés humaines c'est elle qui assure l'évolution et l'adaptation des ces sociétés à travers le temps. Nous ne pouvons imaginer où en serait l'homme s'il réapprenait à chaque génération à tailler des silex ou à contrôler le feu. Mais le problème ne se pose pas car les êtres humains possèdent une faculté d'apprentissage et d'adaptation qui n'est ni prévisible ni mesurable. Il n'en est pas de même pour les entités informatiques. En effet, assurer la transmission du savoir, l'apprentissage et l'adaptabilité des sociétés d'Agents est un véritable sujet qui promet encore de longues années de recherche. Nous tentons d'amener, ici, une petite pierre à cet énorme édifice en proposant un modèle d'apprentissage de connaissances basé sur la communication. Notre travail est le résultat de la mise en commun de deux domaines : l'interprétation des langages et la communication Agent. Notre idée est de profiter de l'apprentissage comme effet secondaire de la communication. En effet, le but de l'éducation est de faire changer d'état son interlocuteur. **Ce changement se fait après l'évaluation (1$^{er}$ domaine) des nouveaux éléments apportés par la communication (2$^{ème}$ domaine)**. Nous proposons dans cet article un modèle permettant de réaliser ce changement. Plus précisément, notre travail est basé sur le modèle STROBE [4], qui considère les Agents comme des interpréteurs Scheme. Ces Agents sont capables d'interpréter les messages d'une conversation dans un environnement donné, incluant un interpréteur, dédié à la conversation courante. Nous allons montrer comment, grâce à la communication, les variables stockées dans ces environnements, et en particulier les interpréteurs, peuvent être modifiés dynamiquement. Ainsi, considérant les environnements d'un Agent comme sa connaissance, nous montrerons comment il apprend plus qu'une simple information, en modifiant, sa façon de voir ces informations et en devenant capable d'en intégrer de nouvelles. Nous illustrerons ce que nous appelons apprentissage au méta-niveau par une expérimentation de dialogue de type « professeur-élève ».

Nous pensons que considérer les Agents comme des interpréteurs est un point de vue très intéressant et nous montrerons comment cela peut être effectif pour des domaines tels que le Web, le GRID, la génération de service dynamique. Ce dernier intérêt constitue le noyau de cet article où nous illustrerons le potentiel d'Agents considérés comme des interpréteurs non déterministes dans un scénario de type e-commerce. Le but de l'article étant de convaincre que sur le Web, l'approche classique du génie logiciel qui consiste à spécifier un problème puis le coder peut être remplacé par une approche qui alterne spécification et exécution basée sur le dialogue. Le but est de réaliser ces deux étapes en même temps, soit de permettre la spécification dynamique d'un problème.

La suite de l'article est organisée de la manière suivante : la partie 2 propose un aperçu des problématiques associées aux notions de communication et d'apprentissage dans les SMA. Nous tentons également de définir un scénario « idéal » pour l'évolution des sociétés d'Agents et pour la communication dans l'avenir. Les parties 3 et 4 présentent notre modèle qui consiste à considérer les Agents comme des interpréteurs Scheme et qui leur fournit un ensemble de couples (environnement interpréteur) pour la représentation des autres. La partie 5 présente brièvement notre architecture Scheme et illustre l'apprentissage au méta-niveau dans un exemple jouet de type «professeur – élève». Puis, la partie 6 illustre la spécification dynamique, après avoir succinctement introduit le concept d'interpréteur non déterministe. Finalement, la partie 7 met en avant les intérêts et extensions de ces principes pour d'autres domaines.

## 2. Communication et apprentissage dans les SMA

La simple mise en commun de plusieurs Agents ne suffit pas à former un SMA (Systèmes Multi-Agents), c'est le fait que ces Agents communiquent qui le permet. C'est grâce à la communication (directe ou non) que sont possibles la coopération et la coordination entre Agent [8]. Définir et modéliser la communication a toujours été difficile. Aujourd'hui, il existe de nombreux modèles et langages de communication mais pouvons nous dire qu'ils sont adaptés au monde Agent ou à de nouvelles formes de communication comme celles que propose le Web ? Il ne s'agit pas de prendre les langages traditionnels de communication ou même les paradigmes actuels de programmation et de les adapter au Web et aux Agents. **Il s'agit de développer de nouvelles architectures et de nouveaux langages conçus pour le Web et les Agents.** En effet, les langages traditionnels sont ficelés et il est souvent très difficile de les faire évoluer. Pour être efficace, la communication doit être intrinsèque à un langage. Par exemple, en ce qui concerne l'apprentissage, il ne suffit pas simplement d'apprendre au niveau *donnée* ou au niveau *contrôle*, il faut aussi apprendre au niveau *interpréteur.*

Nous pouvons faire ressortir quelques pré-requis aux langages de communication Agent : Si nous considérons le fait qu'une communication a des effets sur les interlocuteurs (acte perlocutoire), alors il nous faut obligatoirement considérer que les Agents peuvent changer de but ou de point de vue au milieu de cette communication. Ils doivent donc **être autonomes et s'adapter pendant la communication** [4]. La partie 5 explique brièvement comment notre architecture Scheme nous permet cela. Il faut aussi considérer que des Agents peuvent interagir entre eux ou avec des humains suivant les mêmes principes [12] [6]. Ce qui compte c'est **la représentation qu'un Agent se fait de son interlocuteur**. La partie 6 montrera comment le concept d'Environnement Cognitifs [3] peut nous aider.

Historiquement, les SMA étaient construits avec un langage de communication intégré fonctionnant de manière ad-hoc. Aujourd'hui, la communauté SMA tend à fournir des *Agent Communication Languages* (ACL) applicables à un maximum d'interactions entre Agents. En effet, fournir un ACL fort d'un point de vue sémantique donne un gros avantage pour la création et l'évolution d'un SMA [7]. Ces ACL sont basés sur la théorie des actes de langage[1]. Traditionnellement, les messages KQML ou FIPA-ACL fournissent un élément qui correspond à l'ontologie utilisée dans la communication. Cela permet de rendre les ACL indépendants de n'importe quel vocabulaire et donne à l'interlocuteur un moyen d'établir la correspondance concept / signification des éléments du contenu d'un message. Dorénavant un ACL ad-hoc n'est plus spécifié pour l'incorporer à un SMA, mais une ontologie est construite et passée en paramètre des messages. Nous proposons dans ce papier une alternative à cet état de fait. Les ACL reçoivent souvent la critique du manque de performatif. Notre première expérimentation propose un exemple de solution à ce problème. Elle illustre une technique pour diffuser des nouveaux performatifs dans un SMA, par apprentissage direct d'Agent à Agent.

---
[1] Modèle issu de la philosophie du langage [2] et [15].

Face à ces langages, de nombreux modèles de communication ont été proposés. Une alternative sur la façon de considérer les Agents est présentée dans [11]. Entre autres, STROBE s'intéresse à de nombreux principes importants pour une communication et se base sur les trois primitives Scheme : *STReam, OBject,* et *Environment*. Il met en avant des points comme la représentation de l'interlocuteur, la conservation de l'historique d'une conversation, l'apprentissage issu de la communication, etc. Nous reviendrons souvent aux propositions de STROBE car le modèle proposé ici s'en inspire.

Imaginons maintenant un scénario «idéal» de ce que pourrait être un SMA dans quelques années. Il considère toutes les entités du Web comme des Agents d'une même société qui peuvent communiquer les uns avec les autres naturellement et se transmettre des connaissances. Cette société pouvant s'étendre sans aucune limite. Dans ce SMA, chaque Agent est initialisé avec un minimum requis de connaissances (pour interagir) ainsi qu'avec une spécialité qui le caractérise et qu'il peut transmettre aux autres. Cet Agent possède un ensemble d'interpréteurs qui représentent sa connaissance et son évolution dans le temps. Il apprend tout le reste au fur et à mesure de ses communications. Il apprend même à enseigner ! Pour chaque Agent avec qui il communique, il a une représentation spécifique de celui-ci, ce qui lui permet de tenir compte de ce qu'il apprend tout en gardant éventuellement son comportement et ses croyances d'origine intacts. Bien sur, ces Agents profitent également des méthodes d'apprentissage classique, leur permettant d'analyser leurs environnements locaux (de représentations des autres) pour décider ou pas de modifier leur comportement (leur environnement global). D'un point de vue coopération / coordination, un Agent peut demander à un autre d'interpréter pour lui tel ou tel programme et de lui renvoyer le résultat. Un Agent peut également transmettre à un autre un interpréteur qui lui permet d'effectuer sa tâche. Ces interpréteurs peuvent même être transmis avant une conversation, comme est transmise aujourd'hui une ontologie. Mais vu qu'aucun interpréteur n'évolue de la même manière, puisque deux conversations ne sont jamais identiques, cela donne à cette société d'Agents une pluralité des connaissances incommensurable ! Son évolution devient totalement imprévisible et autonome. Dans cette société toutes les taches sont réalisées par le dialogue. L'intégration des nouveaux Agents se fait naturellement et petit à petit. Il n'est plus possible alors de prouver de manière théorique que tel ou tel Agent sait accomplir une tâche ; le seul moyen est de regarder les solutions émergentes qui apparaissent lorsqu'un problème se pose.

Nous tentons dans cet article de proposer des idées pour rendre réalisable ce scénario encore utopiste aujourd'hui. Entre autres, nous allons voir comment un Agent qui possède plusieurs interpréteurs peut les modifier dynamiquement pour évoluer au fur et à mesure des conversations.

## 3. Les Agents comme des interpréteurs Scheme

Le modèle que nous proposons a pour caractéristique principale le fait qu'il considère les Agents comme des interpréteurs. En s'inspirant de la boucle classique d'évaluation nous considérons les Agents comme des interpréteurs de type REPL (Read, Eval, Print, Listen). Lors d'une communication, chaque Agent exécute une boucle REPL qui sont imbriquées les unes dans les autres. Ce principe est important car il permet de considérer les Agents comme des entités autonomes dont le comportement et les interactions sont dirigées par des procédures concrètes. Notre modèle utilise Scheme aussi bien pour le contenu des messages que pour leur représentation. De cette façon nous pouvons utiliser le même interpréteur pour évaluer le message et son contenu. Exemple d'expression Scheme représentée par des messages :

```
>(define x 2)   ⇔   >(assertion (define x 2))
:x              ⇔   :(ack x)
>x              ⇔   >(request x)
:2              ⇔   :(answer 2)
```

Ce point de vue est très intéressant car les Agents bénéficient de tous les avantages liés à Scheme pour la représentation de connaissances en particulier le modèle de mémoire que constitue l'environnement et le modèle de contrôle obtenu grâce aux procédures et aux continuations de première classe. Par exemple, STROBE propose une structure d'environnement conservant l'historique, basé sur les *streams*, c'est à dire que les liaisons ne sont plus du type (var val) mais du type (var val1 … valn-1 valn). Cette structure devient accessible à des Agents représentés par des interpréteurs.

Pour implémenter ce modèle nous avons écrit un méta-évaluateur Scheme (appelé méta-eval) ainsi qu'un méta-évaluateur non déterministe (appelé méta-ambeval) reconnaissant un certain langage (dans lequel doivent être écrit le contenu des messages d'une conversation) et nous y avons rajouté un module d'interprétation des messages, la fonction `ambevaluate-kqmlmsg`. Ceci nous permet de bien expliciter les trois niveaux d'abstraction : *donnée*, *contrôle*, *interpréteur*. L'apprentissage au niveau *donnée* consiste à affecter des valeurs à des variables déjà existantes ou à définir de nouvelles données (exemple : `(set! a 4)` ou `(define a 3)`) ; Au niveau *contrôle*, consiste à définir de nouvelles fonctions par abstractions sur celles déjà existantes (exemple : `(define (square x) (* x x))`). **Et enfin, l'apprentissage au niveau *interpréteur* ou méta-niveau consiste à faire évoluer directement l'interpréteur correspondant à l'Agent** (exemple : rajouter une forme spéciale). Notre travail propose un mécanisme d'apprentissage à ces trois niveaux et en particulier au troisième. En faisant évoluer son interpréteur[2], un Agent apprend plus qu'une simple information, il change complètement sa façon de percevoir ces informations. C'est la différence entre apprendre une donnée et apprendre à traiter une classe de données.

## 4. Représentation des autres

Un des principes de STROBE est d'avoir un « modèle du partenaire », c'est la notion d'Environnement Cognitif. En effet, pour ce qui est de la représentation de l'interlocuteur, STROBE met en avant le fait qu'il faut avoir un modèle du partenaire pour pouvoir reconstruire son état interne. Ce modèle propose que chaque dialogue soit interprété dans une paire d'environnement : le premier privé, appartenant à l'Agent, et le deuxième représentant le modèle du partenaire courant. Notre travail exploite cette notion. En fait, il se greffe dessus car, comme le concept d'Environnement Cognitif fournit aux Agents un environnement global (ou privé) et plusieurs environnements locaux de représentation des autres, notre proposition fournit aux Agents non pas un interpréteur mais plusieurs interpréteurs dont un global (ou privé) et un pour chaque Agent dont ils ont une représentation. Ainsi, l'évaluation des messages d'une conversation se fait avec un interpréteur donné dans un environnement donné. Notre travail se greffe sur ce concept d'environnement dans le sens où ces interpréteurs, pour être accessibles, doivent eux-mêmes être stockés dans ces environnements. Ainsi nos Agents possèdent les trois attributs suivants :

- **GlobalEnv** leur environnement global.
- **GlobalInter** leur interpréteur global.
- **Other** = {(name, environment, interpreter)} un ensemble de triplet correspondant aux représentations des autres.

Leur environnement global est privé et ne change pas. C'est cet environnement qui est dupliqué[3] lors de l'arrivée d'une nouvelle conversation et c'est son clone, stocké dans un élément de Other, qui est modifié au fur et à mesure de la conversation. La Figure 1 illustre ces représentations.

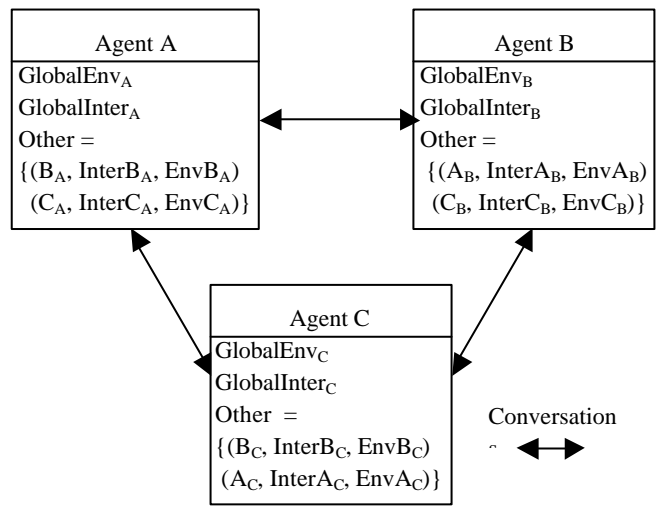

**Figure 1**. *Les attributs d'un Agent et ses représentations des autres.*

## 5. Modification dynamique d'un interpréteur

### 5.1. Architecture Scheme

En Scheme, la boucle classique d'interprétation REPL consiste à attendre que l'utilisateur tape une expression, lire cette expression, l'évaluer renvoyer le résultat et attendre l'entrée de l'expression suivante. Nous sommes alors classiquement au niveau *donnée* et *contrôle*. Le niveau du dessus, *interpréteur*, n'est pas accessible directement. Pour y accéder il faut soit simplement avoir accès aux sources de cet interpréteur (codé dans un certain langage) et modifier ces sources. Soit d'une manière plus sophistiquée si l'interpréteur est réflexif et propose le mécanisme des *reifying procedures*, alors ce mécanisme peut être utilisé pour accéder au contexte d'exécution[4] d'une fonction et le modifier [9]. L'interpréteur est alors dynamiquement modifiable.

Nous utilisons un autre schéma. Au lieu de faire évaluer les expressions utilisateur par l'interpréteur courant (Scheme), nous utilisons celui-ci pour appeler un autre interpréteur (méta-eval), stocké lui dans une structure de donnée (un environnement de première classe que nous contrôlons) qui lui applique le bon interpréteur (méta-ambeval) sur l'expression utilisateur. Ainsi, les expressions utilisateur peuvent consister à modifier l'environnement et en particulier l'interpréteur stocké dans celui-ci. L'interpréteur évaluant l'expression utilisateur est donc accessible et modifiable. Notre idée est de transporter ce schéma dans nos Agents. Ainsi nous pouvons considérer les trois interpréteurs suivants caractérisés par deux procédures :

---

[2] Nous verrons à la partie suivante qu'en fait nos Agents sont vus comme un ensemble d'interpréteur.

[3] Pas forcément si nous considérons un Agent qui voudrait reprendre une conversation dans le contexte d'une autre déjà existante ou ayant existée (avec son environnement et son interpréteur).

[4] Le contexte d'exécution d'une expression est constitué de l'environnement (`r`) d'évaluation de cette expression (`e`) et de la continuation (`k`) qui correspond à la prochaine expression à évaluer avec le résultat.

- Scheme : `eval` et `apply`
- Méta-eval : `evaluate` et `apply-procedure`
- Méta-ambeval : `ambevaluate` et `ambapply-procedure`[5]

Alors la boucle classique d'évaluation est `(eval e r k)`.où e r k est le contexte d'exécution. La notre est `(eval (apply-procedure 'ambevaluate e r ks kf))`.

### 5.2. Protocole de communication

Les messages que nous considérons sont inspirés de la structure des messages KQML ou FIPA-ACL. Ils seront notés par la suite kqmlmsg, ils sont de la forme : **(kqmlmsg performative sender receiver content)**. Pour que méta-ambeval puisse interpréter les messages kqmlmsg, il faut lui ajouter un test qui les reconnaît dans la fonction `ambevaluate`, ainsi qu'une fonction qui les traite en évaluant leur contenu, `ambevaluate-kqmlmsg`. Les performatifs essentiels sont : *assertion, order, ack, executed* ainsi que, nous le verrons, *broadcast*, qui fait l'objet de l'exemple expliqué plus loin. Lorsqu'un Agent reçoit un message indexé par un performatif qu'il ne connaît pas il l'indique de manière particulière :

- Les messages *assertion* ont pour but de modifier le comportement ou les représentations de l'interlocuteur. Leurs réponses sont des messages de type accusé de réception *ack* signifiant un succès ou une erreur.
- Les messages *order* demandent à l'interlocuteur d'exécuter une procédure. Le résultat est envoyé par celui-ci par un message de type *executed*.
- Les messages *broadcast* consistent à envoyer en contenu un couple `(perform, content)` qui signifie que l'interlocuteur doit envoyer un message avec comme performatif `perform` et comme contenu `content` à tous ses interlocuteurs en cours (diffusion). Il n'y a pas de réponse définie pour les messages *broadcast*.

### 5.3. : Exemple d'un dialogue « professeur - élève »

Cet exemple jouet illustre les aspects techniques de notre travail, il montre que ce modèle est viable et peut réellement être implémenté. L'expérimentation montre comment un Agent peut dynamiquement changer sa fonction d'interprétation des messages, donc une partie de son interpréteur. Pour cela nous utilisons l'architecture vue plus haut, chaque Agent possédant un ensemble d'instances de méta-ambeval. C'est un dialogue type « professeur-élève ». Un Agent *teacher* demande à un autre Agent *student* de diffuser (kqmlmsg de type *broadcast*) pour lui un message à tous ses correspondants mais le *student* ne connaît pas le performatif utilisé par le *teacher*. Par conséquent, le

---

[5] Nous avons écrit méta-eval et méta-ambeval en s'inspirant respectivement de [9] et [1].

*teacher* transmet au *student* deux messages (de type *assertion* et *order*) explicitant comment prendre en compte ce performatif. Il lui re-transmet alors son message d'origine et obtient alors satisfaction. Le dialogue exact de l'expérimentation est décrit par la Figure 3.

Nous avons développé pour les besoins de notre expérimentation des Agents capables de communiquer, c'est à dire de s'échanger des messages les uns avec les autres et dy produire des réponses significatives (ceci suivant le protocole définit plus haut)[6]. Ils ne font rien lorsqu'ils ne communiquent pas et leur autonomie se caractérise par le fait qu'ils apprennent seuls. Ils ont comme attributs, name, globalEnv, globalInter, other, ainsi que deux files de messages (une en sortie et une en entrée), et une structure stockant les conversations courantes. Leur comportement consiste uniquement à appliquer la boucle REPL (Figure 2). Notons que ces Agents ne possèdent aucune technique d'apprentissage classique, mais nous nous inscrivons là-dessus dans une logique globale de la communauté IA.

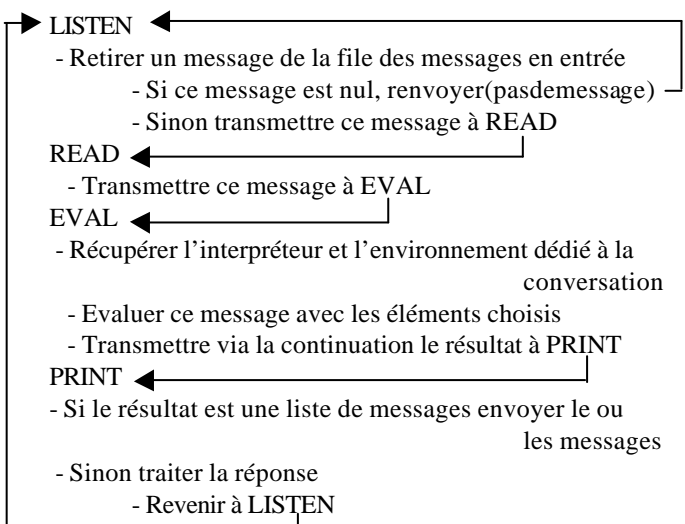

**Figure 2**. *La boucle REPL de nos Agents.*

A l'issue du traitement du dernier message le *student* a modifié sa fonction `ambevaluate-kqmlmsg` et donc sa façon d'interpréter les messages. Le code correspondant à cette fonction dans son environnement dédié à cette conversation est changé. Il est donc maintenant apte à traiter les messages indexés par le performatif *broadcast*.

Remarque (Figure 3) : Le message `learn-broadcast-code-msg` indique comment générer le nouveau code en tenant compte de l'ancienne définition du *student*. Ce code est stocké dans la variable `learn-broadcast-code`.

---

[6] Nos Agents sont en fait des objets programmés en Scheme d'après les techniques présentées par Normark dans *Simulation of Object-Oriented and Mechanisms in Scheme* [13]

| TEACHER | STUDENT |
|---|---|
| Voici la définition de la procédure `square` :<br><br>`(kqmlmsg 'assertion teacher student '(define (square x) (* x x)))` | Ok, je connais maintenant cette procédure :<br><br>`(kqmlmsg 'ack student teacher '(*.*))` |
| Diffuse à tous tes correspondants :<br><br>`(kqmlmsg 'broadcast teacher student '(order (square 3)))` | Désolé, je ne connais pas ce performatif :<br><br>`(kqmlmsg 'answer student teacher ''(no-such-performative broadcast))` |
| Ok, voilà comment ajouter *broadcast* à la liste des performatifs que tu reconnais :<br>Voilà le code que tu devras générer et ajouter à ta procédure `ambevaluate-kqmlmsg` :<br><br>`(kqmlmsg 'assertion teacher student learn-broadcast-code-msg)`<br><br>Exécute cette procédure :<br><br>`(kqmlmsg 'order teacher student '(set! ambevaluate-kqmlmsg learn-broadcast-code)))` | Ok, j'ai rajouté ce code dans une variable de mon environnement :<br><br>`(kqmlmsg 'ack student teacher '(*.*))`<br><br>Ok, je viens de modifier mon évaluateur :<br><br>`(kqmlmsg 'executed student teacher '(*.*))` |
| Diffuse à tous tes correspondants :<br><br>`(kqmlmsg 'broadcast teacher student '(order (square 3)))` | Ok, je diffuse…<br><br>`(kqmlmsg 'order … '(square 3))` |

**Figure 3**. *Dialogue « professeur-élève » pour l'enseignement de broadcast*

## 6. Spécification dynamique par évaluation non déterministe

Avant de regarder comment utiliser des interpréteurs non déterministes pour la génération de service et la spécification dynamique, nous devons introduire la notion de d'interpréteur non déterministe telle qu'elle est présenté dans [1].

### 6.1. Qu'est ce qu'un interpréteur non déterministe ?

L'idée principale est qu'avec un langage non déterministe, une expression peut avoir plusieurs valeurs possibles. Une expression représente en fait un ensemble de « mondes » possibles, chacun déterminé par un ensemble de choix. Un programme peut avoir, en évaluation non déterministe, plusieurs exécutions différentes. L'évaluation non déterministe repose sur la forme spéciale `amb`. L'expression `(amb exp1 exp2 … expn)` retourne une des n expressions `expi`[7]. Par exemple l'expression :

`(list (amb 1 2 3) (amb 'a 'b))` peut avoir 6 valeurs : `(1 a) (1 b) (2 a) (2 b) (3 a) (3 b)`

L'intérêt d'un tel évaluateur est qu'ensuite des fonctions peuvent appeler la forme `amb` en rajoutant des contraintes (sous forme de prédicat) sur les valeurs renvoyées par `amb`. Ces contraintes s'expriment avec la forme spéciale `require` définie comme ceci :

```
(define (require p)
   (if (not p) (amb)))[8]
```

L'évaluation d'une expression `amb` peut être vue comme l'exploration d'un arbre de solutions où le traitement de l'expression continue jusqu'à trouver une solution respectant toutes les contraintes et ceci tant que l'arbre complet n'a pas été parcouru. Lorsque la forme `(amb)` est évaluée cela correspond à une feuille de cet arbre, une autre branche est donc explorée. Pour implémenter un évaluateur non déterministe il faut bien sur gérer la forme spéciale `amb`, et il faut introduire le concept de *failure continuation* (`kf`) qui est la continuation appelé en cas d'échec dans une évaluation. Le contexte d'exécution devient donc `(e r ks kf)` où `ks` correspond à la continuation classique. Un peu comme en Prolog, nous pouvons demander l'ensemble des solutions d'une expression logique une à une avec un évaluateur non déterministe, la forme `try-again` permet de voir la prochaine évaluation à succès d'une fonction appelant `amb`. La boucle classique d'interprétation est modifiée en évaluation non déterministe pour tenir compte de ces retours en arrière (*backtracks*).

---

[7] L'expression `(amb)` sans argument correspond à un échec.

[8] The form `(if cond exp)` returns `exp` value if `cond` is true. Else it return no value.

Considérons la fonction (an-element-of list) qui renvoie la valeur d'un élément d'une liste donnée. Alors son évaluation est :

```
> (an-element-of '(a b c))
: b
> try-again
: a
> try-again
: c
> try-again
: no more values
```

Avant tout, regardons comment est écrit le corps de la fonction an-element-of. La contrainte dans ce cas est que la liste ne soit pas vide. Lorsqu'elle est null? alors (amb) est évaluée et l'évaluateur non déterministe passe à la branche suivante de l'arbre des solutions :

```
(define (an-element-of items)
  (require (not (null? items)))
  (amb (car items)
       (an-element-of (cdr items))))
```

Regardons maintenant un exemple de programme non déterministe un peu plus compliqué tiré de [1]. Il s'agit d'un problème type puzzle logique : *"Baker, Cooper, Fletcher, Miller, and Smith live on different floors of an apartment house that contains only five floors. Baker does not live on the top floor. Cooper does not live on the bottom floor. Fletcher does not live on either the top or the bottom floor. Miller lives on a higher floor than does Cooper. Smith does not live on a floor adjacent to Fletcher's. Fletcher does not live on a floor adjacent to Cooper's. Where does everyone live?"*

Nous pouvons déterminer qui habite à chaque étage en énumérant toutes les possibilités et en leur appliquant les différentes contraintes ; Cela donne la fonction suivante :

```
(define (multiple-dwelling)
  (let ((baker (amb 1 2 3 4 5))
        (cooper (amb 1 2 3 4 5))
        (fletcher (amb 1 2 3 4 5))
        (miller (amb 1 2 3 4 5))
        (smith (amb 1 2 3 4 5)))
    (require (distinct? (list baker cooper
fletcher miller smith)))
    (require (not (= baker 5)))
    (require (not (= cooper 1)))
    (require (not (= fletcher 5)))
    (require (not (= fletcher 1)))
    (require (> miller cooper))
    (require
     (not (= (abs (- smith fletcher)) 1)))
    (require
     (not (= (abs (- fletcher cooper)) 1)))
    (list (list 'baker baker)
          (list 'cooper cooper)
          (list 'fletcher fletcher)
          (list 'miller miller)
          (list 'smith smith))))
```

L'évaluation de cette fonction (multiple-dwelling) renvoie, avec un évaluateur non déterministe : ((baker 3) (cooper 2) (fletcher 4) (miller 5) (smith 1))

### 6.2. Spécification dynamique par la communication dans un scénario e-commerce

Nos Agents sont considérés comme des interpréteurs Scheme donc ils peuvent très bien être vus comme des interpréteurs non déterministes ; C'est à dire reconnaître et traiter les formes amb, require, try-again... Ceci est, en effet, intéressant pour eux car ils pourraient alors résoudre des programmes comme ceux vus dans la section précédente. Mais ce n'est pas le seul intérêt. Pour faire le lien avec notre travail orienté communication Agent, le point le plus intéressant est que nos Agents peuvent construire de tels programmes au fur et à mesure d'un dialogue et appliquer ensuite ces programmes pour donner une réponse ou accomplir une tâche pour un autre Agent. Les contraintes, définissant un programme non déterministe, peuvent être explicitées au fur et à mesure du dialogue en utilisant les outils présentés plus haut, de modification dynamique des fonctions et de la façon dont elles sont interprétées.

Considérons par exemple un dialogue type *e-commerce*, de recherche de billet de train comme celui présenté dans [11]. Un billet est caractérisé par une ville de départ, une ville de destination, un prix, une date. Un Agent *SNCF* est sollicité par un Agent *Client* pour lui faire des propositions de billets. Le dialogue en situation réelle pourrait être :

a. *Client* : Je voudrais un billet de Montpellier à Paris.
b. *SNCF* : Pour quand ?
c. *Client* : Demain avant 10H du matin. Pourriez vous me faire une proposition ?
d. *SNCF* : Voilà, train 34170, départ demain 9H30, Montpellier en direction de Paris, 150€
e. *Client* : Vous n'auriez pas à moins de 100 €?
f. *SNCF* : Voilà, train 34730, départ demain 8H41, Montpellier en direction de Paris, 95€
g. *Client* : Une autre proposition s'il vous plait ?
h. *SNCF* : Voilà, train 34392, départ demain 9H15, Montpellier en direction de Paris, 98€
i. *Client* : Ok, celui-ci me va.

Nous pouvons constater que les interactions **a**, **b**, **c** et **e** consistent à établir les contraintes sur la sélection du billet. Les interactions **d**, **f** et **h** sont des applications de la fonction de recherche de billet avec différentes contraintes. L'interaction **g** correspond à une demande du *Client* pour avoir une autre réponse, soit, pour explorer une autre branche de l'arbre des solutions. La Figure 4 illustre la façon dont ce dialogue peut être traduit en expression Scheme pour être réalisé par nos Agents. L'Agent *Client* transmet ses requêtes sous forme de require et de try-again.

| CLIENT | SNCF |
|---|---|
| *Je voudrais un billet de Montpellier à Paris*<br><br>`(require`<br>`  (eq? depart montpellier))`<br><br>`(require`<br>`  (eq? dest paris))` | Début de la construction de `find-ticket` :<br><br>`(define (find-ticket)`<br>`  (let ((depart(amb *ens-ville*))`<br>`        (dest (amb *ens-ville*))`<br>`        (prix (amb *ens-prix*))`<br>`        (date (amb *ens-date*))))`<br>`    (require`<br>`      (not (eq? depart dest)))`<br>`    (require`<br>`      (eq? depart montpellier))`<br>`    (require`<br>`      (eq? dest paris))`<br>`    (list (list 'depart depart)`<br>`          (list 'destination dest)`<br>`          (list 'prix prix)`<br>`          (list 'date date))))`<br><br>*Pour quand ?* |
| *Demain avant 10H du matin*<br><br>`(require (< date *demain10H*))`<br><br>*Pourriez vous me faire une proposition ?*<br><br>`(find-ticket)` | Modification de la fonction `find-ticket` en lui ajoutant la nouvelle contrainte. Puis exécution de cette fonction.<br><br>`((depart montpellier) (destination paris) (prix 150) (date *dem9H30*))`<br><br>*Voilà, train 34170, départ demain 9H30, Montpellier en direction de Paris, 150€.* |
| *Vous n'auriez pas à moins de 100 € ?*<br><br>`(require (< prix 100))`<br>`(find-ticket)` | idem.<br><br>`((depart montpellier) (destination paris) (prix 95) (date *dem8H41*))`<br><br>*Voilà, train 34730, départ demain 8H41, Montpellier en direction de Paris, 95€.* |
| *Une autre proposition s'il vous plait ?*<br><br>`(try-again)` | Execution de `find-ticket` :<br><br>`((depart montpellier) (destination paris) (prix 98) (date *dem9H15*))`<br><br>*Voilà, train 34392, départ demain 9H15, Montpellier en direction de Paris, 98€* |
| *Ok, celui-ci me va* | |

**Figure 4**. *Dialogue entre l'Agent Client et l'Agent SNCF pour la recherche de billet*

L'Agent *SNCF* commence, au début de la conversation, la construction d'une nouvelle fonction `find-ticket` qu'il modifie et exécute au fur et à mesure de cette conversation. Ces modifications consistent à changer une valeur dans l'environnement de l'Agent *SNCF* représentant l'Agent *Client*. Elles sont réalisées à l'aide des principes présentés précédemment.

Cette idée est très intéressante car c'est le dialogue qui construit le calcul à effectuer et non le contraire. C'est un scénario que nous pourrions retrouver dans de nombreuses applications *e-commerce* ou d'autres du même genre où des Agents doivent construire un programme ensemble pour trouver une solution. L'approche classique de construction d'un programme (la plus fréquemment utilisée dans le génie logiciel), qui consiste à spécifier un programme avant de le coder, est changée par une approche de spécification dynamique pendant la construction d'un programme. C'est-à-dire que la spécification et la réalisation sont faites en même temps.

Ce scénario ressemble et s'adapterait facilement à ceux envisagés sur le GRID [14] [5] où il s'agit de générer des services dynamiquement plutôt que de fournir une prestation prédéfinie et statique.

## 7. Intérêts et extensions de ces principes

L'expérimentation de la 5$^{ème}$ partie montre comment prendre en compte un nouveau performatif, donc comment modifier la fonction d'interprétation des messages d'un Agent. Mais les mêmes principes peuvent être utilisés pour modifier n'importe qu'elle partie de l'interpréteur d'un Agent. Par exemple nous aurions pu faire un exemple qui rajoute `cond` ou `let*` au langage reconnu par notre méta-ambeval. Voir même un Agent qui enseigne à un autre comment transformer son interpréteur en interpréteur paresseux en changeant ses fonctions `ambevaluate` et `ambapply-procedure`. Grâce à ce protocole, nos Agents possèdent en fait un ensemble d'interpréteurs qui représentent leurs connaissances. En effet, ils correspondent aux sous-langages que nos Agents reconnaissent et donc à leurs facultés à effectuer une tâche. Comme vu dans le scénario « idéal », les Agents peuvent effectuer des tâches pour d'autres ou même s'échanger leurs interpréteurs[9] comme dans des architectures GRID où il est plus intéressant de déplacer les processus que les données. Leurs interpréteurs peuvent même être aussi transmis avant une conversation, plutôt que de transférer ontologie.

Imaginons une société d'Agents ou un SMA qui suit ce genre de protocole alors n'importe quel Agent peut apprendre quelque chose d'un autre. Si nous construisons un Agent avec une connaissance et une spécialité le caractérisant alors cette spécialité peut se diffuser au fur et à mesure de ses communications. Pour le Web, ce genre de principe est très intéressant. Considérons un Agent qui joue le rôle de serveur pour une nouvelle application orientée Web et qui utilise une série de performatifs correspondant exactement à ce qu'il doit faire. S'il est construit avec le potentiel pour enseigner ces performatifs alors il s'intégrera très bien à une société d'Agents en les enseignants au début des communications qu'il peut avoir. En outre, pour faire le lien avec XML, nous pouvons considérer une DTD, un XML-Schema et surtout un programme XSL comme des « interpréteurs » de données XML. Le modèle présenté ici permet alors de faire évoluer dynamiquement ces « interpréteurs » et donc les documents XML associés donnant au Web un dynamisme et une adaptabilité inégalable. Cela peut être facilement mis en place considérant l'analogie entre Scheme et XML, vu que les documents XML sont des arbres et quoi de mieux que Scheme pour traiter des arbres. Dans la même idée de nombreux langages liant les Sexpressions (Scheme) et XML apparaissent [10].

---

[9] Ceci s'inspire de l'idée énoncée dans [1] : « *If we wish to discuss some aspect of a proposed modification to Lisp with another member of the Lisp community, we can supply an evaluator that embodies the change. The recipient can then experiment with the new evaluator and send back comments as further modifications.* »

Les principes présentés dans cet article peuvent être également utiles pour d'autres types de scénarios. Imaginons qu'au lieu de transférer la définition de la procédure `square`, l'Agent *teacher* transfère à l'Agent *student* la définition d'une procédure implémentant un algorithme optimisé de résolution d'un problème. Par exemple, la fonction calculant le nombre de Fibonacci en utilisant la mémoization (cette fonction, `memo-fib`, transforme le calcul exponentiel d'un nombre de Fibonacci en un calcul linéaire). Dans ce cas, l'Agent *student*, après avoir appris le performatif *broadcast*, peut jouer le rôle de serveur de grille de calcul (*Grid Computing*) en procédant à une sélection des Agents participants à un lourd calcul utilisant par exemple `memo-fib` de la manière suivante : Il peut demander à tous ses correspondants de réaliser pour lui un calcul simple d'un nombre de Fibonacci et de lui renvoyer le résultat. En fonction de certain critère comme en particulier le temps de réponse cet Agent est capable de sélectionner un ensemble d'Agents et de leur demander de participer à un gros calcul, utilisant les nombres de Fibonacci. Il peut même enseigner aux autres Agents la bonne version de la fonction implémentant cet algorithme. Cette idée est effectivement particulièrement intéressante pour le *Grid Computing* mais peut être également utilisée dans des protocoles de communication type *contract net*.

## Conclusion

Nous avons essayé de montrer dans ce papier une méthode d'apprentissage pour les Agents cognitifs issue de la communication. Cet apprentissage peut se faire par communication simple (niveau *donnée* et *contrôle*), ou par modification interne de l'Agent (niveau *interpréteur*). Nous avons illustré cette idée par un couple d'exemples jouets. Dans le second, nous pouvons voir comment notre modèle convient tout particulièrement à des scénarios de type e-commerce en permettant la spécification dynamique de contraintes par interactions entre Agents.

Si les Agents interprètent de façon dynamique les messages qu'ils reçoivent, ils deviennent adaptables et, sans aucune intervention extérieure, peuvent communiquer avec des entités qu'ils n'ont jamais rencontrées auparavant. En outre, comme leur interpréteur est modifié pour acquérir une connaissance, il peut l'être aussi pour apprendre à enseigner une connaissance. Dans ce cas le transfert du savoir devient exponentiel au fur et à mesure des communications. Ce papier n'a pas simplement pour vocation de proposer un artéfact de plus de programmation à ajouter aux Agents, mais l'idée est plutôt de montrer une technique, faite de manière simple et utilisable, d'évolution autonome des Agents dans une société.

## Annexe

Le modèle présenté ici a été sujet à une petite implémentation, non compète encore à ce jour, mais d'ores et déjà fonctionnelle. Elle est disponible en ligne sur http://www.lirmm.fr/~jonquet. Elle fut réalisée avec MIT Scheme 7.7.1, norme R$^5$RS. Vous y trouverez les expérimentations citées ici.

## Remerciements

Ce travail a été réalisé en grande partie pendant le stage de DEA d'un des auteurs (CJ). Le support du projet européen LeGE-WG (Learning Grid Excellence Working Group) est amplement remercié.